# Reconfigurable and Recyclable Low-Threshold Quasi-BIC Lasers via a Tunable polymer Coating

Xiaolin Wang#, Jiayao Liu#, Zimeng Zeng, Hongyu Yuan, Zhuoyang Li, Zelong He, Wenping Gong, and Zhaona Wang*

Key Laboratory of Multiscale Spin Physics (Ministry of Education), Applied Optics Beijing Area Major Laboratory, School of Physics and Astronomy, Beijing Normal University, Beijing 100875, China

*Contact author: zhnwang@bnu.edu.cn

**Abstract**

Reconfigurable and sustainable microcavity lasers are highly desirable for next-generation integrated photonics. Here, we report a recyclable, low-threshold quasi-bound state in the continuum (q-BIC) laser fabricated via low-cost, high-throughput interference lithography. By introducing a polyvinyl alcohol (PVA) coating on a dye-doped photonic crystal, we suppress out-of-plane symmetry breaking, which reinforces optical confinement and reduces the lasing threshold. The q-BIC modes are further tuned through tailoring the refractive-index of the PVA layer by using Kramers-Kronig (K-K) relation via Rhodamine 6G doping, demonstrating a wavelength shift of 7.14 nm and a sensitivity of 215 nm RIU$^{-1}$ as a sensing prob. More importantly, lasing modes are reversibly tuning via precisely controlling the coating thickness. Exploiting the dissolving and re-coating process, the laser is repeatedly reconfigured while maintaining performance. This work provides a sustainable and adaptive platform for sensing and reconfigurable photonic systems.

**Introduction**

Micro- and nanocavity lasers have emerged as essential building blocks for next-generation photonic technologies owing to their compact size, ease of integration and capability for coherent light generation. These miniature coherent light sources hold great promise for diverse applications in integrated photonic circuits, sensing, and laser display technologies[1-9]. Over the past decades, various mechanisms have been developed to achieve low threshold lasing[10-15]. Among them, optical bound states in the continuum (BICs)[16-18], which support perfectly confined modes despite their spectral overlap with radiating waves, have been recognized as an effective means to enhance light confinement and dramatically increase the quality (Q) factor in periodic photonic structures[19-20]. The BIC-based lasers have gained remarkable progress, including reduced thresholds, narrow linewidths, and directional emission[21-23]. However, most reported devices rely on high-refractive-index dielectric materials such as GaAs[24], $TiO_2$[25], or $Si_3N_4$[26], which require high-precision nanofabrication techniques including electron-beam lithography and reactive-ion etching. These fabrication processes are costly, time-consuming, and incompatible with large-area or flexible substrates. Furthermore, the Q factor of experimentally realized periodic structures is inevitably constrained by fabrication imperfections and finite sample size. As a result, the ideal BIC state with infinite Q factor is transformed into a quasi-BIC (q-BIC) mode with a finite Q factor, leading to increased lasing thresholds and degraded performance[6, 8]. To overcome the limitations associated with high-cost and small-area fabrication, interference lithography has been looked as a powerful alternative for producing large-area periodic structures with high uniformity and tunable periodicity[27]. Using this technique, various lasing systems have been demonstrated, such as perovskite-based photonic arrays fabricated by spin-coating photoresist layers followed by interference exposure[28], or by directly patterning dye-doped photoresist films to form large-area gratings[29]. However, such structures are typically exposed to air, possessing a large refractive index contrast between the substrate and the surrounding medium. This vertical asymmetry enhances out-of-plane radiation and reduces the Q factor, thereby increasing the lasing threshold. To suppress these radiative losses, it is crucial to develop an effective approach to restore out-of-plane symmetry while maintaining large-area scalability.

Another critical limitation of current q-BIC lasers lies in their limited tunability. In most cases, the lasing wavelength is typically determined by fixed geometric parameters such as the grating period or slab thickness[28-29]. To achieve their tunability, several external stimuli strategies have been proposed as post-fabrication methods. Thermal modulation of gain media has been used to obtain wavelength shifts of around 10 nm between 273 K and 383 K[30]. The electric-field control has also been demonstrated to tune wavelength up to 28 nm via the electro-optic effect[31]. Alternatively, environmental refractive-index modulation has been proposed as a sensing with a theoretical sensitivity of exceeding 200 nm RIU$^{-1}$[32]. Nevertheless, these methods often involve irreversible processes, high energy consumption, or complex device configurations. Moreover, these methods remain at the proof-of-concept stage. Therefore, it remains highly desirable to develop a low-cost, reversible, and scalable approach for dynamically tuning the lasing characteristics of BIC-based systems.

Here, we present a reconfigurable and recyclable q-BIC laser platform based on a polymer-covered grating architecture, enabling both low-threshold operation and reversible wavelength tunability through tailoring the vertical symmetry and additional waveguiding role. The device is fabricated via a simple, high-throughput two-beam interference lithography process, followed by the deposition of a polyvinyl alcohol (PVA) overlayer. The introduction of the PVA film effectively restores out-of-plane symmetry by reducing the refractive index contrast between the substrate and the cover layer, thereby enhancing optical confinement and lowering the lasing threshold from 0.47 to 0.37 mJ cm$^{-2}$ under the transverse electric (TE) wave pump. Furthermore, the PVA overlayer enables post-fabrication tuning of the lasing wavelength by tailoring the BIC modes through modulation of optical parameters such as thickness and refractive index (RI). Via the Kramers-Kronig (K-K) relations, the lasing wavelength is precisely controlled, yielding a RI sensor with an experimental sensitivity of 215 nm RIU$^{-1}$. Specifically, fine-tuning the overlayer thickness allowed for the control of radiative laser modes, leading to the observation of single-mode lasing from a quasi-BIC at a specific thickness. Furthermore, the cyclic application of dissolving and spin-coating processes enabled the reversible switching of q-BIC lasing modes and the repeated reuse of the grating structure. This polymer-assisted strategy offers a facile, cost-effective, and sustainable route to reconfigurable q-BIC lasers, thereby providing a promising platform for uncovering the rich band characteristics and proposing scalable and adaptive photonic integration.

**Results and discussions**

The design concept and structural characterization of the proposed q-BIC laser are presented in Figure 1. As shown in Figure 1a, the traditional fabricated grating structure possesses inherent vertical asymmetry due to the RI difference ($\Delta n$) between the substrate ($n_s$) and the air cladding. This asymmetry results in substantial radiation loss and a low Q-factor for the optical modes. Such high optical loss impedes lasing action by necessitating a high threshold. To address this issue of vertical asymmetry, a PVA layer was employed on top of the dielectric grating to reduce the RI contrast $\Delta n$ between $n_s$ and the coating layer ($n_c$), which effectively enlarges the $Q$ value of the q-BIC mode. To elucidate the effect of the PVA overlayer on the optical resonances, the photonic band structures of the bare and PVA-coated grating were calculated using the finite element method (FEM) and shown for the TE ($y$-polarized) modes in Figure 1b and transverse magnetic (TM) modes in Figure S3 (Supporting Information). The observed optical resonances originate from the coupling of Bloch resonances characterized by the Bloch wavevector $\boldsymbol{\beta}_n = (k_x + n\frac{2\pi}{a})\boldsymbol{u}_x$ [33-34], denoted as TE$_{m,n}$ and TM$_{m,n}$, where $m$ represents the waveguide mode index and $n$ corresponds to the diffraction order along the $x$-direction. Modes with the same $m$ value couple at the Γ point, forming symmetry-protected BICs, while coupling between modes with different $m$ values at off-Γ angles gives rise to Friedrich-Wintgen q-BICs (FW-qBICs) through $z$-direction symmetry breaking[25, 35]. For the TE waves, the bare grating gives a distinct bandgap at approximately 576.7 nm. A symmetry-protected BIC named as BIC$_1$ is identified at the Γ point of band I within this gap, exhibiting the characteristic antisymmetric electric field distribution along the $x$-direction (Figure 1c1)[36]. Upon the introduction of the

PVA coating, the BIC modes redshifts due to the increased local RI and additional symmetry-protected BICs appears at the interface layer. At a thickness of $h_1$=2400 nm, two such symmetry-protected BICs emerge at the Γ point of bands II and III, named as $BIC_2$ (580.7 nm) and $BIC_3$ (560.3 nm) with their antisymmetric electric field distributions in Figure 1c2 and 1c3. Notably, the optical field of the symmetry-protected $BIC_2$ mode is primarily confined within the waveguide layer beneath the grating, while that of the $BIC_3$ mode is mostly localized near the interface between the grating and the PVA overlayer. Figure 1d summarizes the *Q*-factor dispersion of the three bands corresponding to the three symmetry-protected BICs, where the *Q* factors diverge to infinity at Γ point, confirming their non-radiative nature. The PVA-coated structure achieves a significantly higher *Q* factor than its bare counterpart, owing to better RI matching with the glass substrate, even with a narrower bandgap. Moreover, the anti-phase distribution of the electric field along the *z*-direction at the $BIC_3$ further suppresses radiation loss via destructive interference, leading to a maximal *Q* factor among the three symmetry-protected BICs. To better understand the decreased radiation loss in enhancing *Q* factor via PVA coating layer, a Hamiltonian model is constructed based on coupled mode theory (Details in Supporting Information). The interaction between Bloch modes $TE_{m, +1}$ and $TE_{m, -1}$ is considered to explain the emergence of BIC modes at the Γ point. The photonic band structures and *Q* factor distributions for the bare and PVA-coated grating structure (Figure S1a and 1b, Supporting Information) are calculated using this analytical model, which indicates the suppress radiation loss role of the PVA coating as the above analysis. Simultaneously, two high-*Q* FW-qBICs were observed at an oblique angle of 1.7°in Figure S2 (Supporting Information) with almost asymmetry electric field along the *z*-direction. For the TM waves, the PVA coating layer plays a similar role in suppressing radiation loss and increasing the localization of modes (Figure S3, Supporting Information). These findings demonstrate that the optimized optical environment effectively minimizes radiation leakage and enhances optical confinement, thereby supporting high-*Q* q-BIC resonances that are favorable for achieving low-threshold lasing.

In our experiments, a complementary grating architecture composed of a periodic grating and a PVA overlayer is achieved for the q-BIC lasing device as shown in Figure 1e. The fabrication process began with the preparation of a bare grating on a glass substrate via two-beam interference lithography (Figure S4, Supporting Information), using a photoresist film doped with Pyrromethene 567 (PM567) as the gain medium. The choice of PM567 was motivated by its photoluminescence (PL) emission band (550–650 nm, Figure S5, Supporting Information) meeting the targeted q-BIC lasing modes. A grating structure was then achieved through precise control of the exposure and development parameters, with its configuration characterized by top- and side-view scanning electron microscope (SEM) images (Figure 1f). Finally, a thin PVA overlayer was spin-coated onto this dye-doped grating, forming the PVA-coated structure. The thicknesses of the PVA overlayer and the underlying waveguide layer are $h_1$=2400 nm and $h_2$=480 nm, while the ridge height and width of the grating are $h_3$=700 nm and $w$=170 nm, respectively. Angle-resolved reflection spectra of the bare and PVA-coated gratings were acquired by a home-built microspectroscopy system (Figure S6, Supporting Information) which supports reflection and PL signal collection. For the bare grating structure (Figure 1g1), the low-energy band progressively narrows and vanishes as the angle of incidence approaches zero at 574 nm [22] under continuous-wave TE-polarized illumination, agreeing with the characteristics of BIC. In contrast, the PVA-coated structure exhibits weakened intermodal coupling accompanied by the increased number of localized modes, leading to a possibility to control lasing modes. Similarly, for TM polarization, corresponding BICs are clearly observed in both the bare and PVA-coated grating configurations (Figure S7, Supporting Information).

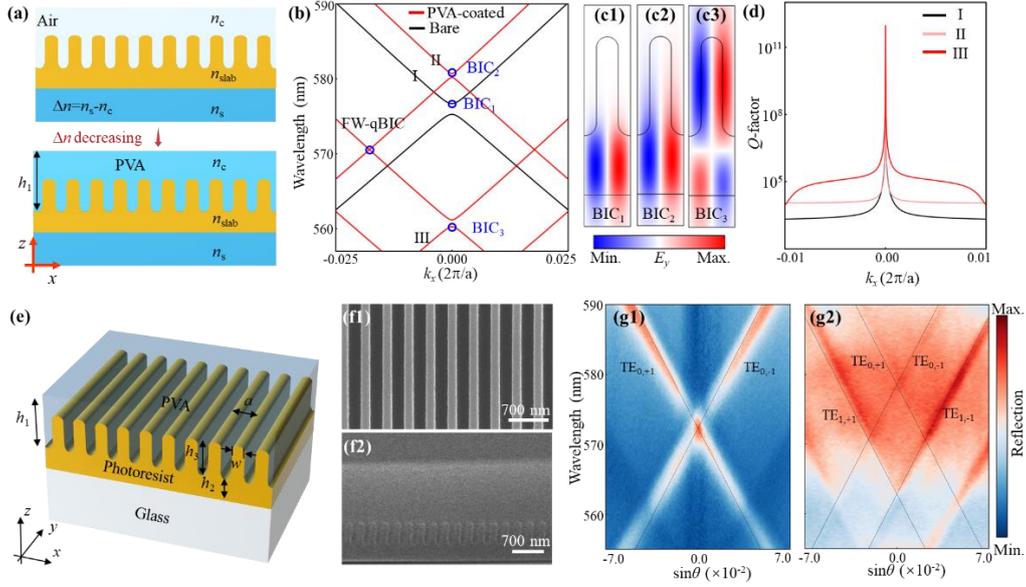

**Figure 1.** Design concept and characterization of the proposed q-BIC laser device. (a) Schematic illustration of the design concept, showing that the introduction of a PVA overlayer reduces the index contrast ($\Delta n$) between the substrate ($n_s$) and the coating layer ($n_c$), thereby enhancing the $Q$ factor of the q-BIC mode. (b) Simulated band structures for the bare (black lines) and PVA-coated (red lines) grating structure. (c) The distributions of electric field component $E_y$ of the $BIC_1$ (c1), $BIC_2$ (c2) and $BIC_3$ (c3) modes at $\Gamma$ point. (d) $Q$-factors dispersion of bands I (black line), II (light-red line) and III (red line) identified in panel (b). (e) Schematic of the designed q-BIC laser with a complementary grating configuration. (f) Top-view (f1) and side-view (f2) SEM images of the fabricated sample. Scale bars: 700 nm. (g) Experimental angle-resolved reflection spectrum of the bare (g1) and PVA-coated (g2) grating structure. Black lines indicate the uncoupled original Bloch modes.

The lasing performances of the designed q-BIC resonators were systematically characterized using angle-resolved PL spectroscopy under the excitation of a 532 nm nanosecond pulsed laser (1 kHz repetition rate, 0.8 ns pulse width). As shown in Figure 2a, the angle-resolved PL spectrum of the bare grating under TE polarization reveals two distinct Bloch modes of $TE_{0,\pm1}$, agreeing with the reflection results in Figure 1g1. Once the pump fluence exceeds the threshold, a sharp emission peak appears at 574.2 nm. The emergence of a characteristic two-lobe far-field profile near the $\Gamma$ point also confirms the q-BIC lasing action associated with $BIC_1$. In contrast, the PVA-coated grating exhibits more complex modal behavior (Figure 2b). In this structure, four Bloch modes of $TE_{0,\pm1}$ and $TE_{1,\pm1}$ are clearly resolved in the angle-resolved PL spectra, consistent with the reflection features in Figure 1g2. When pump energy is above the threshold, two strong emission peaks are observed at the Γ point. One is at 560.6 nm, corresponding to a q-BIC mode near $BIC_3$, while another mode is at 578.2 nm, associated with $BIC_2$. Additionally, two weaker lasing modes appears at off-Γ angles, which are regarded as the FW-qBICs. The distinct lasing behaviors of the two structures are further demonstrated by the PL spectra for different TE-polarized pump energies (Figure 2c). For the bare grating (Figure 2c1), a broad spontaneous emission background is firstly observed with pump energy below 0.42 mJ cm$^{-2}$. As the pump energy increases to 0.47 mJ cm$^{-2}$, a sharp peak emerges at 574.2 nm, accompanied by a rapid increase in intensity and significant linewidth narrowing, confirming the single-mode lasing behavior from the q-BIC mode near $BIC_1$. In contrast, the PVA-coated grating exhibits more complex lasing behavior (Figure 2c2). A sharp lasing peak appears at 560.6 nm when pump energy density increases to 0.37 mJ cm$^{-2}$, attributed to the q-BIC mode near $BIC_3$. With continuously increasing pump energy, the second peak emerges at 578.2 nm ($BIC_2$), followed by the third mode at 569.3 nm above 0.42 mJ cm$^{-2}$, originating from FW-qBIC at oblique angles.

The lasing performance was further proved by the integrated photoluminescence (PL) intensity and spectral linewidth against pump energy (Figure 2d). All modes exhibit a characteristic "S-shaped" intensity growth accompanied by a pronounced linewidth collapse, confirming the transition from spontaneous to stimulated emission. The lasing thresholds were determined to be ~0.47 mJ cm$^{-2}$ for the 574.2 nm mode in the bare grating (black curve), and ~0.37 mJ cm$^{-2}$ and ~0.42 mJ cm$^{-2}$ for the 560.6 nm ($BIC_3$, red) and 578.2 nm ($BIC_2$, light red) modes in the PVA-coated device, respectively. The markedly lower thresholds in the coated structure align well with the theoretically predicted enhancement in optical confinement. As designed, the q-BIC mode near $BIC_3$ exhibits the lowest threshold among all modes. Interestingly, despite its early onset and minimal threshold, this mode still shows limited intensity growth and is progressively suppressed after the emergence of the $BIC_2$-related mode. It is because that its smaller gain volume leads to a negative impact when intermode competition occurs.

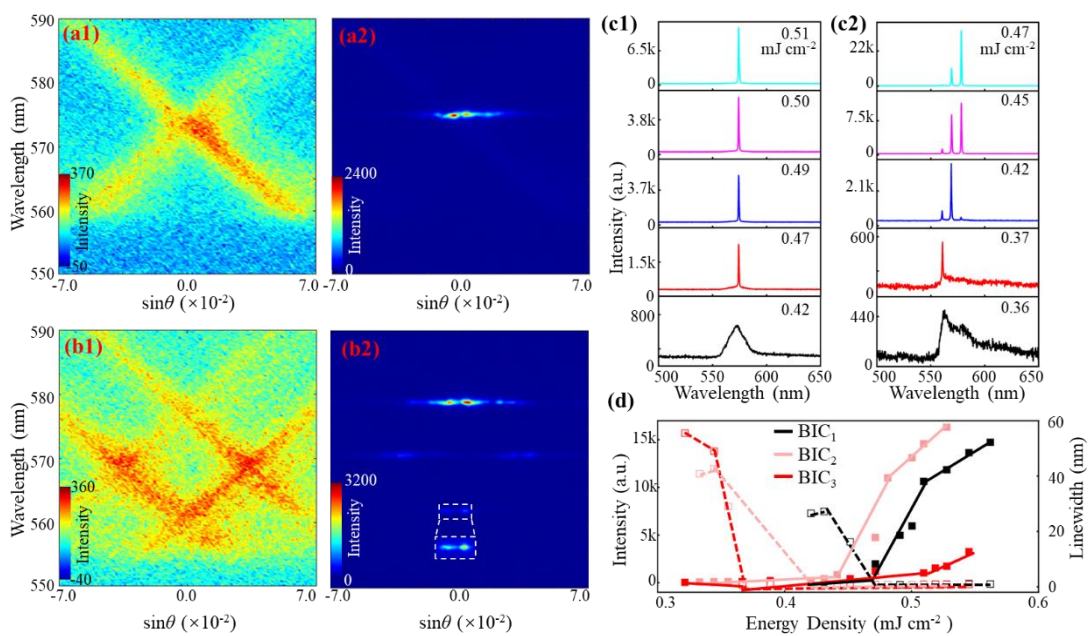

**Figure 2.** Lasing performance of the designed q-BIC lasers. (a, b) Angle-resolved PL spectra of the bare grating(a) and the PVA-coated grating (b), measured below (a1, b1) and above (a2, b2) the lasing threshold, respectively. A magnified view of the white dashed-boxed region is presented in the inset. (c) Emission spectra of the bare (c1) and PVA-coated (c2) devices under different pump fluences. (d) Pump fluences-dependent spectral integrated intensities (open squares) and FWHMs (squares) of the lasing modes. The black data set corresponding to the mode at 574.2 nm in the bare grating with a lasing threshold of 0.47 mJ cm$^{-2}$, while the red and light red data sets corresponding to modes at 560.6 nm and 578.2 nm in the PVA-coated grating with thresholds of 0.37 mJ cm$^{-2}$ and 0.42 mJ cm$^{-2}$, respectively.

When it comes to TM polarization, both the bare grating and the PVA-coated grating maintain the typical two-lobe far-field radiation pattern characteristic of q-BICs in the angle-resolved PL spectra (Figure S8, Supporting Information). Consistent with the TE-polarized case, the PVA-coated grating also shows a lower lasing threshold (0.38 mJ cm$^{-2}$) when compared with the bare grating (0.53 mJ cm$^{-2}$). These findings verify that the PVA overlayer effectively strengthens optical confinement and reduces radiation loss, thereby improving lasing efficiency under both TE and TM polarizations.

Given the analogous modulation effects of the PVA overlayer on both TE and TM waves, we focus our subsequent dynamic control studies on the TE polarization.

To effectively utilize the localized energy field in the PVA overlayer, rhodamine 6G (R6G) molecules were introduced into the PVA matrix. R6G was selected due to its excellent water solubility and a PL spectrum (550-610 nm, Figure S5, Supporting Information) that provides sufficient gain for the coating-layer modes. In addition, the inclusion of R6G allows to modulate of the effective RI of the coating, thereby enabling fine adjustment of the quasi-BIC mode. As the R6G concentration increases, the PVA film exhibits an enhanced absorption (Figure S9, Supporting Information). According to the Kramers-Kronig (K-K) relations [37-39], the enhanced absorption leads to a corresponding increase in the real part of the RI. The RI of the dye-doped PVA films with varying R6G concentrations was experimentally detected by spectroscopic ellipsometry. Figure 3a summarizes the extracted RI values as a function of R6G concentration, with insets showing optical photographs of films doped with 1.0, 3.0, and 5.0 mg mL$^{-1}$ doping. The progressively deepening color also indicates the increasing dye concentration. According to figure 3a, the real part of RI gradually increases when R6G concentration climbing, indicates the effectiveness of RI tailoring. Since the experimentally observed lasing modes are centered near 560 nm, the corresponding RI values at $\lambda$=560 nm were used for subsequent simulations. To investigate the influence of RI variation on the device eigenmodes, we simulated normal-incidence TE-polarized reflectance spectra using rigorous coupled-wave analysis (RCWA) as shown in Figure 3b. All resonance wavelengths exhibit a pronounced redshift with increasing RI, suggesting the strong potential of the proposed q-BIC laser for RI sensing.

Experimentally, lasing spectra from devices with varying R6G concentrations are presented in Figure 3c. As the concentration increases, the lasing wavelengths progressively redshift, and multimode lasing emerges. Angle-resolved reflection and PL spectra (Figure S10, Supporting Information) reveal that the shortest-wavelength lasing mode corresponds to the couple of the TE$_{3, +1}$ and TE$_{3, -1}$ modes at the $\Gamma$ point. This mode is strongly localized within the PVA overlayer, whereas the additional lasing peaks at longer wavelengths originate from the coupling of TE$_{m, \pm 1}$ modes (with different $m$ values) at oblique angles. Focusing on the shortest-wavelength lasing mode, the simulated and experimental RI sensitivities were determined as $S_{sim}$=246 nm RIU$^{-1}$ (Figure S11, Supporting Information) and $S_{exp}$=215 nm RIU$^{-1}$ (Figure 3d), respectively, using $S = \Delta\lambda/\Delta n$ [40-41]. The discrepancy between $S_{sim}$ and $S_{exp}$ arises from the neglected optical dispersion of the dye-doped PVA film in simulations. Notably, the experimentally measured sensitivity is comparable to the theoretically predicted performance of q-BIC-based RI sensors[32], highlighting the strong potential of the designed q-BIC laser platform for optical sensing applications.

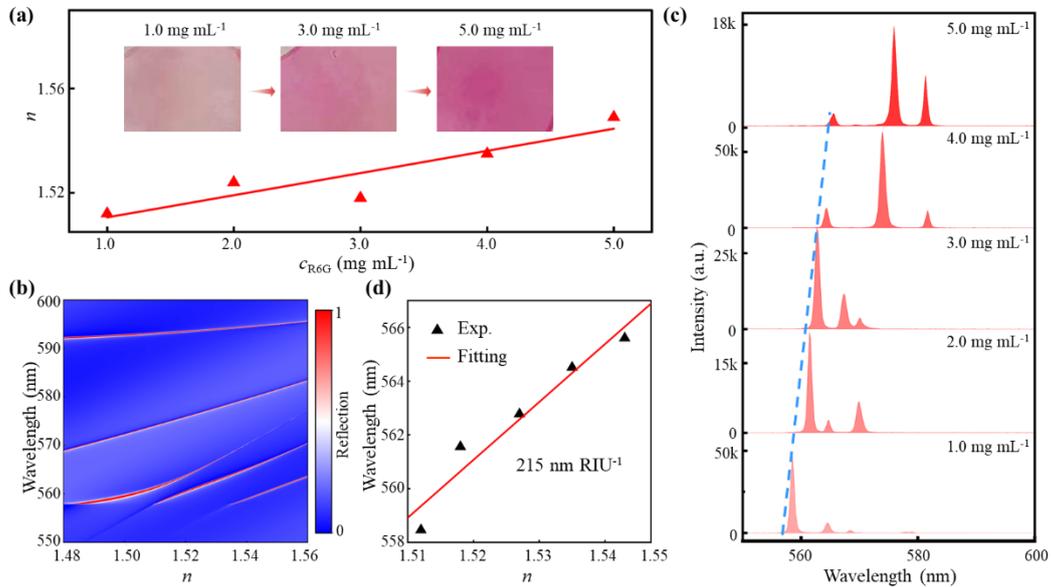

**Figure 3.** Performance of designed q-BIC laser in refractive-index sensing. (a) RI values of dye-doped PVA films as a function of R6G concentrations, with optical images of representative samples shown in the inset. (b) Simulated reflectance spectra under normal-incidence TE-polarized illumination as a function of overlayer RI, obtained via RCWA. (c) Experimental lasing spectra of the samples with various R6G concentrations. (d) Linear relationship between experimental resonance wavelength and RI of the overlayer.

Beyond RI control, we exploit an alternative strategy for tailoring q-BIC lasing by precisely modulating the thickness of PVA overlayer. Thanks to the water solubility of PVA, we are able to thin the overlayer by controlling water-dissolution cycles, as illustrated in Figure 4a. To systematically unravel the influence of thickness on the modes, we simulated normal-incidence TE-polarized reflectance spectra using RCWA across a range of PVA thicknesses (Figure 4b). The simulations reveal that as the PVA layer thins, all guided-mode resonances undergo a gradual blueshift, accompanied by a reduction in the number of supported modes. The structure eventually transits into a single-mode regime when the thickness is reduced below ∼100 nm. Experimentally, we tracked this evolution through angle-resolved measurements. After eight water-dissolution cycles, the reflection spectrum (Figure 4c1) resolves eight distinct Bloch modes, assigned to $TE_{0,\pm1}$, $TE_{1,\pm1}$, $TE_{2,\pm1}$ and $TE_{3,\pm1}$. The corresponding lasing spectrum (Figure 4c2) reveals a q-BIC mode near the Γ point, originating from the coupling of the $TE_{3,\pm1}$ pair. After fourteen dissolution cycles, the reflection spectrum (Figure 4d1) shows only six remaining modes—$TE_{0,\pm1}$, $TE_{1,\pm1}$ and $TE_{2,\pm1}$—which display a clear blueshift relative to their counterparts in the thicker film. Consistently, the lasing spectrum (Figure 4d2) at this stage features a q-BIC mode generated by the coupling of $TE_{2,\pm1}$ pair.

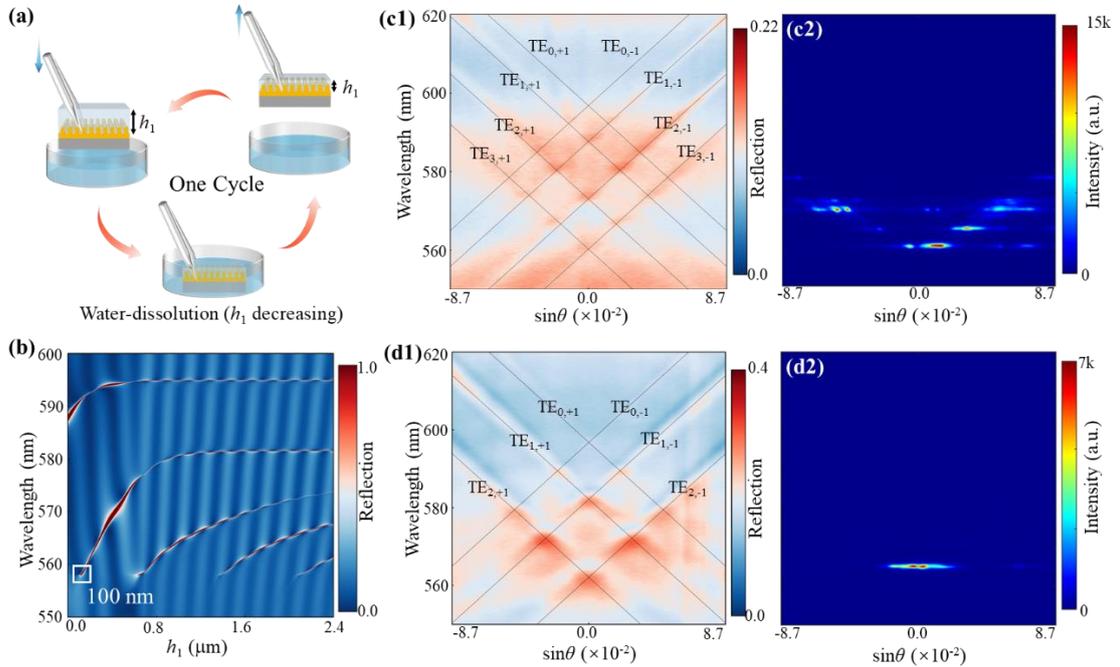

**Figure 4.** The q-BIC lasing mode control by the thickness of the overlayer. (a) Schematic illustration of the water-dissolution cycles used to modulate the thickness of the PVA overlayer. (b) Simulated reflectance spectra under normal-incidence TE-

polarized illumination as a function of the PVA thickness using RCWA. (c) Experimental angle-resolved reflection (c1) and PL (c2) spectra of the sample after eight water-dissolution cycles demonstrates a multi-mode lasing behavior. (d) Experimental angle-resolved reflection (d1) and PL (d2) spectra of the sample after fourteen water-dissolution cycles demonstrates a single-mode lasing behavior.

We further monitored the dynamic spectral evolution under continuous dissolution via lasing emission measurements (Figure 5a) and angle-resolved PL spectroscopy (Figure 5b). As dissolution progresses, all lasing modes exhibit a continuous blueshift, confirming that thinner overlayers support shorter resonant wavelengths. With increasing cycles, originally observed modes vanish and newly emerging modes continue to blueshift. By sixteen cycles, a pronounced spectral reorganization yields a lasing peak near 580 nm that aligns with bare grating emission, indicating the nearly complete removal of PVA layer. The emission pattern at this stage varies from multimode competition to a single dominant mode. Collectively, these results demonstrate that water-dissolution enables continuous and accurate tailoring of the PVA overlayer thickness. This geometric parameter directly governs the photonic band structure, as evidenced by the systematic blueshift of resonant modes and the reduction in mode count with decreasing thickness. This establishes thickness modulation as a robust and reversible route for dynamically tailoring q-BIC lasing characteristics.

Capitalizing on the precise thickness-tuning of the PVA overlayer, we demonstrate a fully recyclable q-BIC laser through the washing-recoating cycles. As schematically illustrated in Figure 5c, this regenerable platform enables the device to be repeatedly reconfigured among PVA-coated, bare, and recoated states via controlled dissolution and redeposition processes. Crucially, recoating with a fresh PVA layer restores the original lasing characteristics, completing a full reversible reconstruction cycle. To assess the practical reliability of this approach, we systematically evaluated the laser performance over multiple reconstruction iterations. Emission wavelength and peak intensity remain highly stable across devices subjected to ten and sixteen dissolution cycles (Figure 5d). Furthermore, optical photographs (Figure S12, Supporting Information) verifies that the grating structure retains its integrity throughout repeated recycling without signs of deformation or degradation. These findings verify that water-mediated PVA-coated grating configuration provides a robust, reversible, and reproducible strategy for reconfigurable q-BIC lasers, establishing a practical pathway toward reusable and tunable photonic devices.

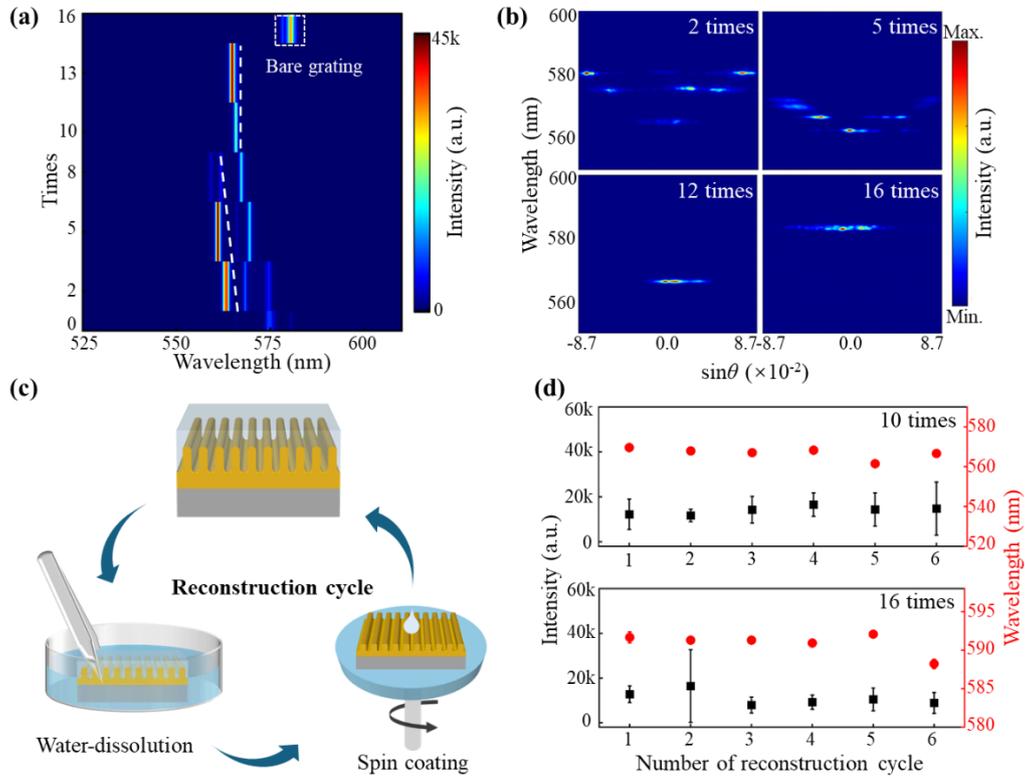

**Figure 5.** Reconfigurable and reusable q-BIC laser enabled by the water-soluble PVA overlayer. (a, b) Lasing spectra (a) and angle-resolved lasing spectra (b) of the sample subjected to different numbers of dissolution cycles. (c) Schematic illustration of the cyclic process for device reconfiguration. The laser device can be reconstructed through successive water-dissolution and spin-coating steps. The PVA overlayer can be completely removed by controlling the number of water-dissolution cycles. (d) Intensity and wavelength of the q-BIC laser as a function of repeated washing-restructuring cycles, recorded from the 10$^{th}$ (top) and 16$^{th}$ (bottom) washing steps, demonstrating a high stability of the lasing performance across multiple reuse processes.

**Conclusions**

We have developed a reconfigurable and reusable q-BIC laser based on a complementary grating structure fabricated via a low-cost, high-throughput two-beam interference lithography. By introducing a water-soluble PVA overlayer onto the one-dimensional grating, we constructed a PVA-coated grating architecture that effectively enhances RI matching with the glass substrate, significantly suppresses radiation leakage, and improves the optical quality factor, thereby reducing the lasing threshold from 0.47 mJ cm$^{-2}$ for the bare grating to 0.37 mJ cm$^{-2}$ for the complementary grating. Importantly, this strategy circumvents the need for complex micro/nanofabrication or etching processes, enabling efficient quasi-BIC lasing at low cost. Through doping R6G molecules, we further modulated the effective RI, experimentally verifying the K-K relation and achieving refractive-index-sensitive lasing with a sensitivity of S = 215 nm RIU$^{-1}$. Leveraging the water solubility of PVA, we also demonstrated continuous thickness tuning via controlled dissolution cycles, allowing precise spectral shifting and mode-number control of the lasing emission. The fully reversible " washing-recoating" process exhibits excellent structural integrity and emission reproducibility, confirming the device's reconfigurability and reusability. This work establishes a simple, sustainable, and material-efficient strategy for dynamically tailoring quasi-BIC resonances and radiation properties, overcoming the limitations of functional rigidity and material waste in conventional nanophotonic systems. It also opens a pathway toward recyclable photonic lasers and future intelligent laser platforms by incorporating functional polymer overlayers responsive to optical, thermal, or chemical stimuli.

## Methods

**Numerical Simulations.** The photonic band structures, Q-factors, and electric field distributions were simulated using the finite element method (FEM). The $Q$-factor was calculated as $Q=(\text{Re}[f])/(2\text{Im}[f])$, where Re[$f$] and Im[$f$] denote the real and imaginary parts of the eigenfrequency of a cavity mode, respectively. Periodic boundary conditions were applied along the $x$-direction, and perfectly matched layers (PMLs) were implemented along the $z$-direction to simulate an open domain. Angle-resolved reflection spectra were computed using rigorous coupled-wave analysis (RCWA) to investigate the optical response under various structural parameters. The refractive indices of the glass substrate and the photoresist were set to 1.50 and 1.63, respectively.

**Sample Fabrication.** One-dimensional PVA-coated q-BIC laser devices were fabricated through a two-step process involving the patterning of dye-doped photoresist gratings followed by deposition of a PVA overlayer. A PM567-doped photoresist solution with a concentration of 2.0 mg mL$^{-1}$ was prepared by dissolving 10 mg of PM567 in 5 mL of photoresist (AZ® MiR™ 701 Series). Separately, a 20 wt.% PVA aqueous solution was obtained by dissolving PVA (Sigma-Aldrich) in deionized water. R6G-doped PVA solutions with concentrations of 1.0, 2.0, 3.0, 4.0, and 5.0 mg mL$^{-1}$ were prepared by dispersing 5, 10, 15, 20, and 25 mg of R6G in 5 mL of the 20 wt.% PVA solution, respectively. The PM567-doped photoresist film was spin-coated onto a cleaned glass substrate at 600 rpm for 10 s followed by 1500 rpm for 60 s, yielding a film thickness of approximately 1.18 μm. The gratings were then patterned using a home-built two-beam interference lithography setup, with a He-Cd laser ($\lambda$ = 325 nm) as the exposure source. After development in AZ 300MIF developer, large-area grating structures were obtained as the bear grating. Finally, R6G-doped PVA solutions with various concentrations were spin-coated onto the gratings at 800 rpm for 10 s followed by 1800 rpm for 60 s, producing uniform overlayers with a thickness of approximately 2.40 μm.

**Optical Characterization.** All optical measurements were performed using a custom angle-resolved imaging and spectroscopy system (Figure S6, Supporting Information). A nanosecond pulsed laser (MCA-532-1-60-01-PD, $\lambda$=532 nm, repetition rate=500 Hz) served as the excitation source. The pump power was adjusted using a combination of linear polarizers and a half-wave plate. Angle-resolved reflection and photoluminescence (PL) spectra were collected using a spectrometer (Andor SR-500i-D1-R), and momentum-space images were acquired with a CCD camera.

**Thickness Control of PVA Overlayer**. The thickness of the PVA coating was precisely regulated by adjusting the number of water-dissolution cycles, with the immersion duration in each cycle carefully controlled to achieve fine thickness modulation. Complete removal of the PVA layer was achieved after 16 cycles, which was verified by comparing the luminescence spectrum with that of the bare grating.

**Cycling Stability Evaluation**. To evaluate the reconfiguration stability, the sample was subjected to repeated washing-reconstruction cycles. Each complete cycle consisted of a 16-step water-dissolution process followed by a reconstruction step. The procedure was as follows: the device first underwent 10 dissolution cycles, after which it was rapidly dried and optically characterized. It was then subjected to six additional dissolution cycles to completely remove the PVA layer, with optical characterization confirming the recovery of the bare grating structure. Finally, a full reconstruction was achieved by re-spin-coating the R6G-doped PVA solution. This "dissolution-measurement- dissolution-recoating" protocol was repeated multiple times. The lasing peak wavelength and intensity, recorded after the 10th and 16th dissolution cycles, were used to systematically assess the performance stability and reproducibility.

## Supporting Information

Supporting Information is available from the Wiley Online Library or from the author.

## Acknowledgements

This work was supported by the National Natural Science Foundation of China (Grant Nos. 92150109 and 61975018) and Beijing Key Laboratory of High-Entropy Energy materials and Devices，Beijing Institute of Nanoenergy and Nanosystems（No. GS2025ZD011）.

## Conflict of Interest

The authors declare no conflict of interest.

## Data Availability Statement

The data that support the findings of this study are available from the corresponding author upon reasonable request.